\documentclass[runningheads]{llncs}
\usepackage{graphicx}
\usepackage[frozencache=true,cachedir=minted-cache]{minted}

\begin{document}
\title{Introducing Traceability in GitHub for Medical Software Development
}
%
%
\author{Vlad Stirbu\inst{1} \and 
Tommi Mikkonen\inst{2,3} 
}
\authorrunning{V. Stirbu et al.}
%
\institute{CompliancePal, Tampere, Finland\\ 
\email{vlad.stirbu@compliancepal.eu}
\and
University of Helsinki, Helsinki, Finland\\
\email{tommi.mikkonen@helsinki.fi}
\and
University of Jyväskylä, Jyväskylä, Finland\\
\email{tommi.j.mikkonen@jyu.fi}
}
\maketitle              
\begin{abstract}
Assuring traceability from requirements to implementation is a key element when developing safety critical software systems. Traditionally, this traceability is ensured by a waterfall-like process, where phases follow each other, and tracing between different phases can be managed. However, new software development paradigms, such as continuous software engineering and DevOps, which encourage a steady stream of new features, committed by developers in a seemingly uncontrolled fashion in terms of former phasing, challenge this view. In this paper, we introduce our approach that adds traceability capabilities to GitHub, so that the developers can act like they normally do in GitHub context but produce the documentation needed by the regulatory purposes in the process. 

\keywords{Traceability  \and Regulated software \and Continuous software engineering \and DevOps \and GitHub.}
\end{abstract}

\section{Introduction}

Assuring traceability from requirements to implementation is a key element when developing safety critical software systems. Traditionally, this traceability is ensured by a waterfall-like process, where phases follow each other, and tracing between different phases can be managed with relative ease. To support this tracing, sophisticated software systems have been implemented, which take advantage of this phasing and help developers to focus on issues at hand in the current phase. 

However, new software development paradigms, such as continuous software engineering \cite{fitzgerald2017continuous} and DevOps \cite{lwakatare2019devops}, which encourage a steady stream of new features, committed by developers in a seemingly uncontrolled fashion in terms of former phasing, challenge this view. Instead of advancing in phases from specification to design to development in the same pace with all features, developers can select items from specification to work on, and eventually they commit new code back to the main codebase. This code is then automatically deployed to use, leaving virtually no trace between specification and the code, unless special actions are taken by the developers.

In this paper, we propose introducing traceability features to GitHub, the most popular site used by software developers. With these features, the developers can act like they normally do while developing software in GitHub context, but also produce the documentation needed by the regulators in the process. A prototype implementation has been built, following the ideas proposed in \cite{stirbu2020compliancepal} as future work. The work has been carried out in medical context, but we trust that the same approach can be applied in other safety critical application domains covered by regulations. However, in the rest of this paper, we focus on the medical domain, as regulatory restrictions may vary across the domains.

The rest of this paper is structured as follows. In Section 2, we present the background and motivation of this work. In Section 3, we address the concept of design control, which is an essential part of designing software intensive medical products. In Section 4, we introduce the proposed approach, relying largely on GitHub concepts. In Section 5, we discuss our key observations and propose some directions for future work in connection with the proposed approach. Finally, we draw the conclusions in Section 6.

\section{Background and motivation}



Medical device software development has unique needs. Its design, development, and manufacturing processes are strictly regulated. To comply with these regulations, there must be proper control mechanisms in place to ensure the end product’s safety, reliability, and ability to meet user needs. These control mechanisms originate from the regulations’ requirements, corresponding guidance documents, international standards, and national legislation. However, their plentiful existence is one of the reasons medical software is often considered a complex domain by developers.

In more detail, for every phase within the product lifecycle -- design, development, manufacturing, risk management, maintenance, and post-market processes -- certain standards must be followed for regulatory compliance. The set of applicable standards for software include general requirements for health software product safety (IEC 82304-1) \cite{iec82304}, software life cycle process (IEC 62304 \cite{iec62304}), risk management process (ISO 14971 \cite{iso14971}), and usability engineering (IEC 62366-1 \cite{iec62366-1}). Furthermore, the manufacturers are expected to have a quality management system that must comply with further associated regulations --  requirements of the Medical Device Quality Systems standard ISO 13485 \cite{iso13485} or its US counterpart, US FDA 21 CFR part 820. These standards form a minimum yet an overwhelming set of regulations to consider when developing medical devices with software.

To ensure compliance to the above standards, plan-driven methodologies have been the preferred way to develop products in regulated industries. Their cultural affinity with the language and format used by standards referred to above have made them the natural choice. However, the long feedback loops that characterize these methodologies are even longer in the high ceremony process required to comply with regulations. Furthermore, these practices are often somewhat distant from development activities that are used in non-regulated software development. Sometimes Application Lifecycle Management (ALM) tools, commonly used in regulated development, amplify this distance rather than helping to overcome it. 

The situation becomes particularly complex when working with medical systems that consist of software only. The developers may have no experience at all in regulated activities, and, once the development activities are initiated, they should have adequate knowledge in regulation-related tasks as a part of the development. Although, the legally binding legislation texts and international standards describe the expected results, they do not describe how to achieve those results. Therefore, practical expertise is required to define the steps required to achieve the objectives \cite{laukkarinen2017devops}. To complicate matters further, many of the available ALM tools require that the developers invest time and effort to keep them in sync instead of relying on automation. 

To deal with the situation, software developers -- who are professionals in software development, not regulation -- often resort to compliance over-engineering or adding extra effort to compliance-related activities to play it safe. This sometimes results in a view that compliance as the necessary evil that must be considered but has little practical relevance. Consequently, the compliance activities are often put aside while creating software and resurrected only when a new feature development task is completed. This resurrection often needs support from dedicated compliance personnel, which might not be fluent with the latest development methodologies.


The developers are not all wrong. The benefits of agile methods and continuous software engineering also apply to medical software. Still, using them in medical software development introduces the same concerns as with any technology -- how to deal with legal and regulatory bindings in a new context \cite{wagner1999keepers}. This culminates in the context of continuous software development, where new releases can be made several times a day, but this is not leveraged because of regulatory constraints. Instead, the developers are stopped from deploying things until all the compliance and regulatory related processes are complete, breaking the natural flow of the development team.  

To complicate matters further, regulatory affairs professionals have often practiced in environments where the medical devices always include hardware, and where they typically follow linear development model. Hence they might not have the skills and experience to operate in an agile software development environment, in particular when medical devices that only include software are considered.

\section{Design control in software intensive medical products}

\begin{figure}[t]
\centering
\includegraphics[width=0.60\textwidth]{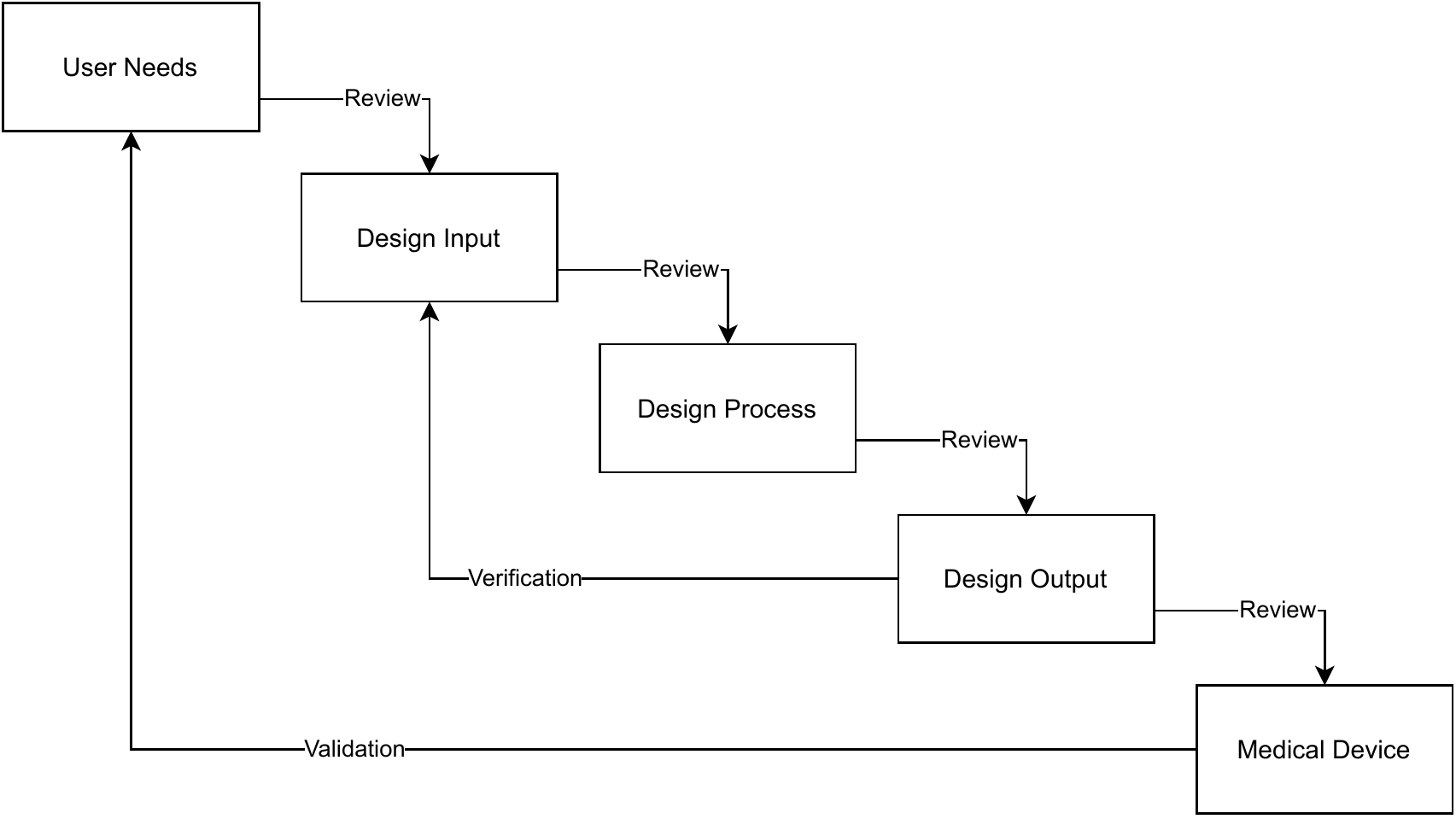}
\caption{Application of design controls to waterfall design process \cite{fda-design-control}} 
\label{fig:design-control}
\end{figure}

The concept of design control is a key element of a quality management system, which ensures that the manufacturer is able to deliver products that fulfill the user needs. The manufacturer is able to ensure, via systematic reviews, that the identified user needs are transformed into actionable design inputs that can be used in a design process to  obtain the design output, which serves as the medical device. Besides the reviews, the manufacturer needs to perform specific activities that ensures that the design output verifies the design input, and that the resulting medical device validates the user needs, as illustrated in Fig. \ref{fig:design-control}.

\begin{figure}
\centering
\includegraphics[width=0.95\textwidth]{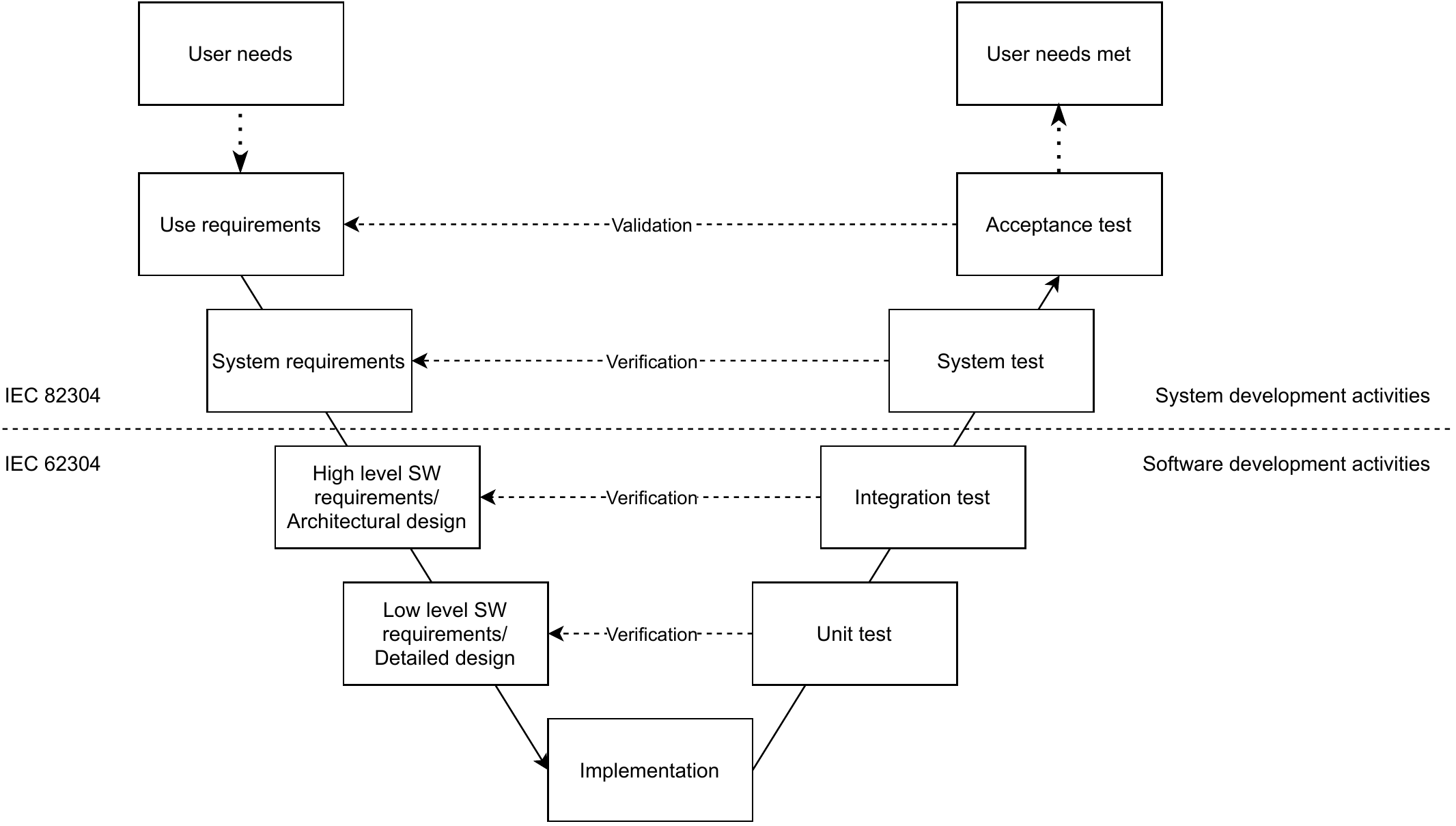}
\caption{System and software development design control activities} 
\label{fig:sdlc-activities}
\end{figure}

For software intensive medical products the design control activities can be split into two layers, depicted in Fig. \ref{fig:sdlc-activities}: the product and system development activities (IEC 82304 \cite{iec82304}), and the software development activities (IEC 62304 \cite{iec62304}). At the product level, the identified user needs are converted to system requirements that serve as design inputs for the the software development process. During software development, the system requirements are transformed into high level software requirements that cover the software system and architectural concerns. Later on, the high level software requirements are further distilled into low level software requirements that serve as design input for implementation.

The resulting code, test cases and various other artifacts, such as architecture and detailed module design documentation, created during the software development activities, serve as the design outputs. The review of the artifacts and the automated test result provide an effective verification procedure at unit, integration and system level. Automated acceptance tests together with the result reports of clinical trials serve as the validation procedure. All these procedures ensure that the proper design controls have been applied during development, resulting in a medical product that meets the user needs.

The design control activities mentioned in IEC 82304 and IEC 62304 are intended to describe only the required activities and desired outcomes, but not the practical ways to achieve them. This approach gives the medical device manufacturers the leeway that allows them to customise their quality management system and software development methodology to reach the intended results. However, it is up to the manufacturers to ensure that the defined quality management system and methodology are compliant to the regulatory requirements.

\section{Proposed approach}
\label{sec:our-approach}

In the following, we describe our approach for implementing effective design controls and collect traceability artifacts using the GitHub native capabilities. First, we describe the information model used for implementing the traceability. We continue with an overview of the GitHub capabilities that serve as enablers of traceability infrastructure. Then, based on a  prototype implementation, we describe how we mapped the information model into the GitHub context, and how we automated the traceability process using GitHub actions.

\subsection{Traceability information model}

To be effective for a software intensive product, the design controls and the traceability audit trail have to be applied to the concepts and tools that are used by the development team during their daily activities. In this context, a team developing medical product using an agile software development methodology and DevOps practices would be familiar with concepts like requirements that cover high level concepts such as user stories, or fine grained details of an implementation. They would be refining the user stories into implementation specifications during the iteration planning, would implement the requirements, and would integrate the product increment after the successful iteration review.

\begin{figure}[t]
\centering
\includegraphics[width=0.75\textwidth]{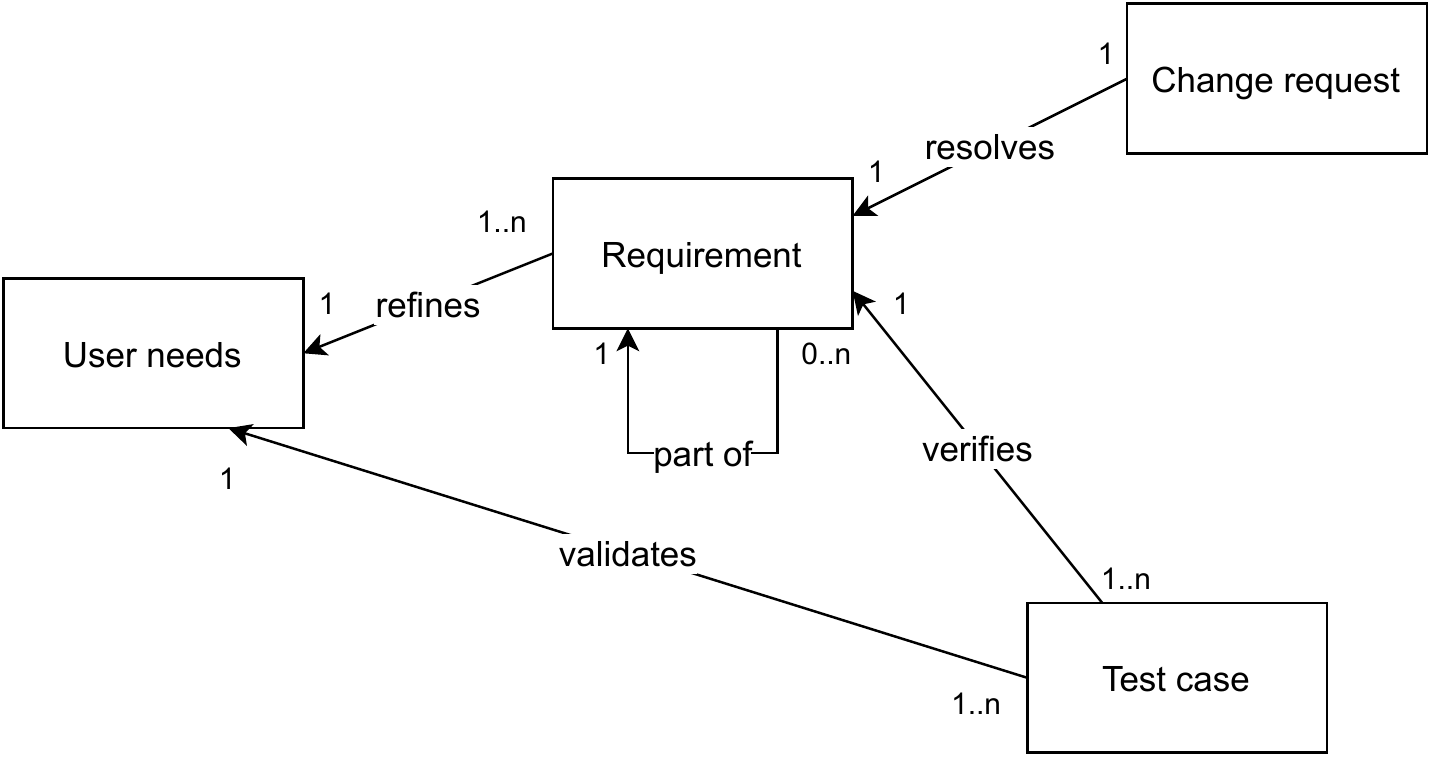}
\caption{Traceability information model} 
\label{fig:information-model}
\end{figure}

Our approach leverages this situation and builds an information model around  \textit{user needs}. The user needs are \textit{refined} into system requirements, that are further \textit{decomposed} into high level and low level software requirements. Each user need can validated by one or more acceptance \textit{test case}. Similarly, a requirement can be verified using a relevant test suite at unit, integration or system level, matching the corresponding requirement scope. The user needs, requirements and test cases serve as design inputs. The implementation of a requirement is modeled as a single \textit{change request}. The change request bundles the code changes, configurations needed to build and run the iteration in scope, automated and manual test results, as well as design artifacts that describe the architecture and detailed implementation of a module. Together, the contents of the change request represents the design output. The change request becomes part of the product after it is verified in a formal review. The entities and the links between them convey in an effective manner the design control and the evidence in the form of an audit trail. The resulting traceability information model is depicted in Fig. \ref{fig:information-model}.

\subsection{Native GitHub enablers}

Over the years, GitHub has expanded their offerings with features beyond git. In the following, we provide a brief overview of the capabilities leveraged for design control and traceability in our prototype implementation.

\subsubsection{Issues}

Every GitHub hosted repository has an Issue section that enables teams to document and track the progress of requirements, specifications of work items, software bugs, feedback from users relevant for the scope of the software developed in the respective repository. An issue has a short title and a body that contains the detailed description using markdown\footnote{https://github.github.com/gfm/}. The body can include \textit{references} to other issues in the same or in a different repository. The references build semantic links between various issues, that can be traversed using the web user interface. Besides the title and the body that contains the description, the issue has associated metadata like \textit{labels}, which allows categorization of issues, and \textit{assignees}, which allows tracking who is performing the work.

\subsubsection{Pull Requests}

GitHub flow is a lightweight branching model that allows teams to work on several work items simultaneous. With this model, the workflow starts with a branch that is created from the code main branch. As the feature is developed the changes are committed to the branch. When the feature implementation is considered complete, the \textit{pull request} is opened signaling the intent to merge into the main branch. Opening the pull request marks the beginning of the \textit{review} phase, during which the assigned members of the team discuss the changes created by the implementation, and fix any problems that are identified. To facilitate the review process, GitHub runs automated test and include the results in the pull request metadata. When the review is complete the feature is merged and becomes part of the product. Linking a pull request with the corresponding issues that describes the feature is achieved by using keywords followed by the reference in the pull request description, e.g. \texttt{resolves \#10}.

\subsubsection{Actions}

GitHub makes easy to automate the software development workflows with \textit{actions}. Although the actions are typically used for automating the building, testing and deploying steps of a software development process, they can be used for other purposes due to their ability to run custom jobs in response to any GitHub event, or even third party events. As such, actions are an effective way to extend the functionality of GitHub and enforce custom workflows, relieving team members from doing repetitive compliance related jobs that can be done better with automation.

\subsection{Prototype implementation}

The prototype implementation relies on the GitHub native capabilities described above. The key features of the prototype are introduced below.

\subsubsection{Mapping to GitHub native capabilities}

As a first step in implementing the design controls and traceability audit trail, we need to map the information model to the capabilities available in GitHub. The use needs, the system and software requirements are implemented as issues labelled with the following labels: need, system requirement and software requirement. The issue creation in the correct format is facilitated by issue templates, which relieves the creator from the chores of ensuring that the issue structure (e.g. sections) and labels are fulfilled. The change requests are implemented with pull requests, while the structure of the pull request is enforced using the pull request template. The relations between issues are implemented using references. Finally, the test cases are described using Gherkin syntax\footnote{https://cucumber.io/docs/gherkin/reference/} or Robot Framework\footnote{https://robotframework.org}. The mapping is summarised in Table \ref{table:mapping}.

\begin{table}[t]
\centering
\caption{Mapping traceability to GitHub native capabilities}\label{table:mapping}
\begin{tabular}{l|l|p{6cm}}
\hline
Traceability & GitHub capability & Implementation\\
\hline
user need & issue & user need template \\
system requirement & issue & system requirement template \\
software requirement & issue & software requirement template \\
change request & pull request & pull request template \\
relations & references & reference to related concepts in issues and pull requests body\\
test case & - & gherkin or robot framework\\
\hline
\end{tabular}
\end{table}

\begin{listing}
\begin{minted}[fontsize=\footnotesize]{md}
## Issue section

Section description

---
partOf: #6

---
\end{minted}
\caption{Issue body source with requirement relationship metadata}
\label{listing:parent}
\end{listing}

\begin{figure}[h!]
\centering
\includegraphics[width=0.75\textwidth]{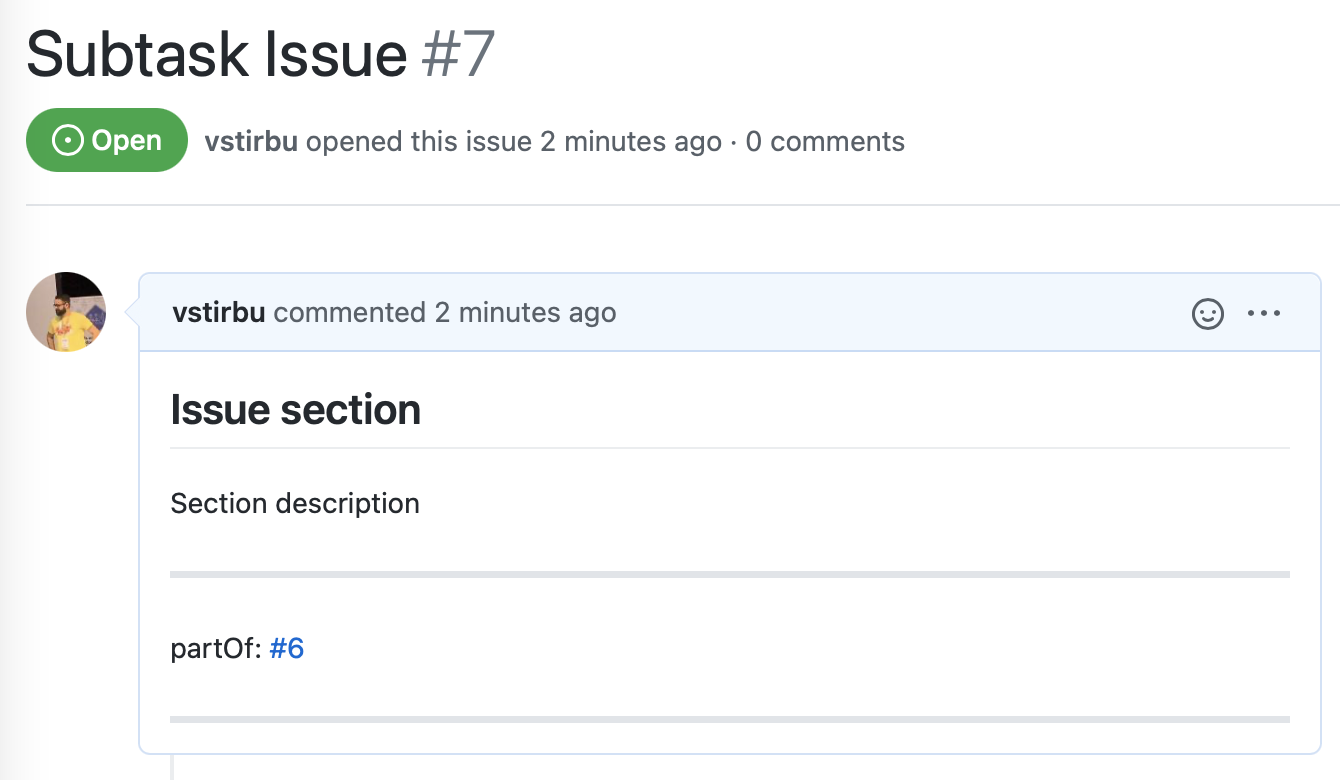}
\caption{GitHub rendering of an issue containing requirement relationship metadata} 
\label{fig:sub-requirement}
\end{figure}

\subsubsection{Conveying parent requirement relationships}

While GitHub is capable of encoding relationships between the issues, it lacks the ability to add semantics to the relationship. In our implementation, we decided to add the semantic information using the frontmatter, a YAML\footnote{https://yaml.org/spec/1.2/spec.html} formatted object that encodes issue metadata, typically located at the beginning or the end of the issue's description. The parent issue is indicated using \texttt{partOf} metadata. In the issue body presented in Listing \ref{listing:parent}, the parent of the issue is the issue \#6 in the same repository. The issue is rendered by GitHub as seen in Fig. \ref{fig:sub-requirement}.

\subsubsection{Visualizing related sub-requirements}

\begin{figure}[h!]
\centering
\includegraphics[width=0.75\textwidth]{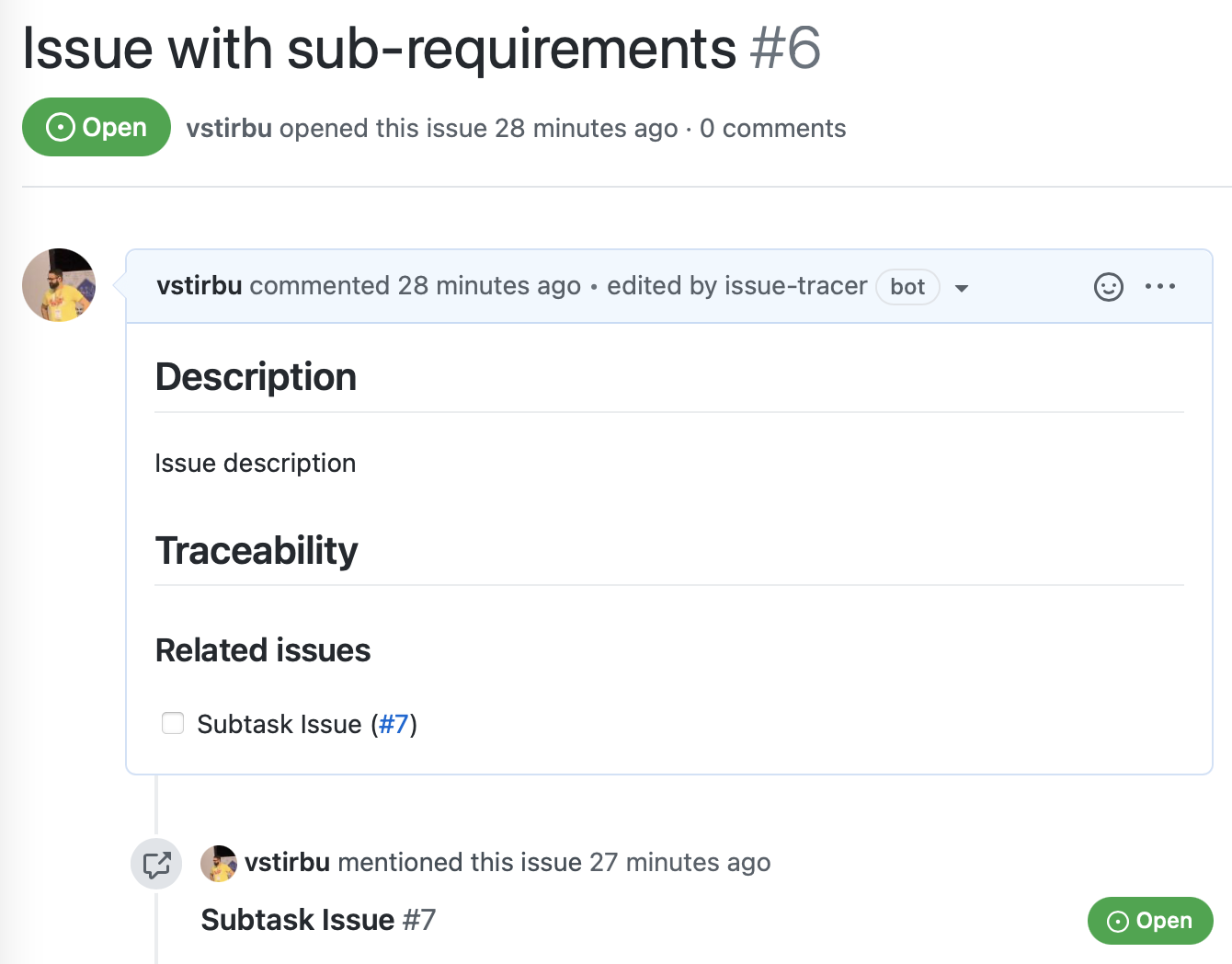}
\caption{GitHub rendering of an issue containing sub-requirements} 
\label{fig:parent-req}
\end{figure}

To better visualize the issues that have been refined in sub-requirements, we are using the ability of GitHub to render markdown checklists. In Listing \ref{listing:sub-requirements}, we can see that the issue \#7 defined earlier is listed as a related issue in its parent issue \#6. We can also encode the status (e.g. open or closed), depending on the state of the corresponding checklist item. The GitHub rendering of this issue is depicted in Fig. \ref{fig:parent-req}.

\begin{listing}[t]
\begin{minted}[fontsize=\footnotesize]{md}
## Description

Issue description

## Traceability

### Related issues

- [ ] Subtask Issue (#7)
\end{minted}
\caption{Issue body source with sub-requirements encoded as a checklist}
\label{listing:sub-requirements}
\end{listing}

\subsubsection{Linking change request with requirements and test cases}

\begin{figure}[h!]
\centering
\includegraphics[width=0.75\textwidth]{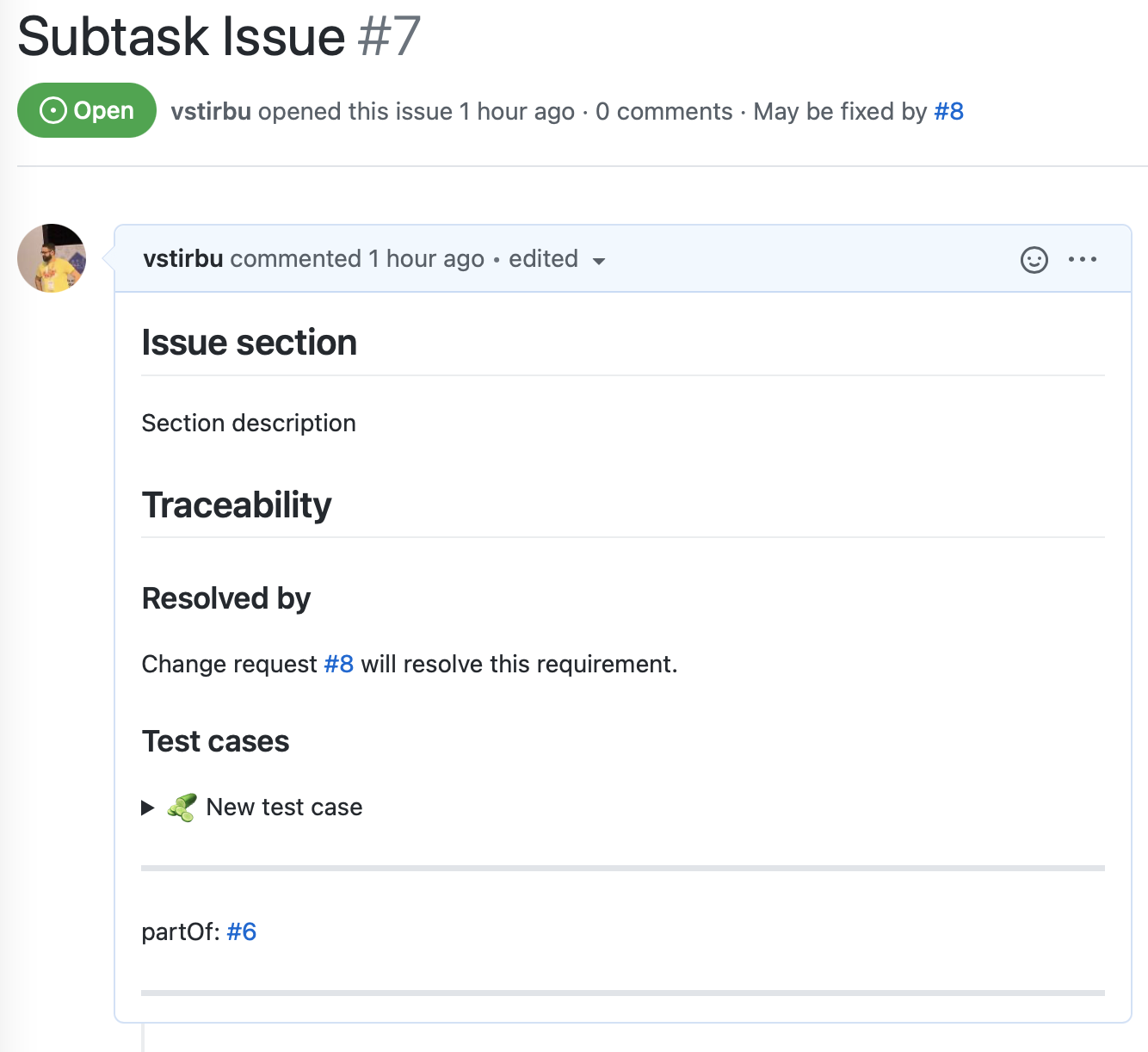}
\caption{GitHub rendering of an issue resolved by a change request and the associated test cases} 
\label{fig:pr-test}
\end{figure}

GitHub has a built-in ability to link pull requests with issues using keywords such as \texttt{Resolves} followed by a reference to the corresponding issue. The capability goes further, as when an pull request is merged the linked issue is automatically closed. Our prototype implementation leverages this capability for building the traceability audit trail between the change request with the requirement resolved by it. In addition we construct relationships between the new test cases introduced by the pull request and the requirement. For example, the test case described in Listing \ref{listing:test-case}, indicates that the new scenario tagged with \texttt{@issue-7} corresponds to requirement \#7. The information is included an the \textit{Traceability} section of the issue and rendered by GitHub as seen in Fig. \ref{fig:pr-test}.

\begin{listing}[t]
\begin{minted}{gherkin}
  @issue-7
  Scenario: New test case
    Given initial state
    When the trigger
    Then resulting state
\end{minted}
\caption{Test case described using Gherkin syntax}
\label{listing:test-case}
\end{listing}

\subsubsection{Automation with GitHub actions}

GitHub user interface is able to render the descriptions of the issues, enabling the users to see the traceability information and traverse the link relations. However, crafting by hand the markdown according to the conventions used in this prototype implementation is laborious and prone to errors. 

To overcome this obstacle, we have automated the process using GitHub actions. Our custom action reacts to \textit{issue} and \textit{pull\_request} events as follows. When an issue event is triggered, the action inspects the body of the issue looking for parent relationship. If found, the action updates the parent issue with information about sub-requirements. Similarly, when the pull\_request event is received, the action detects which issue the change resolves and updates the corresponding information about the test cases. When an issue is merged the status change is reflected in the issue by GitHub and out action updates the status in the parent requirement. As a result, the process of crafting the issue descriptions is performed mostly automated, leaving only two steps in which the user input is needed to indicate the parent relationship and the new issue.


\section{Discussion}

Based on the experiences with the prototype, we next consider two key goals of this work. These address the effectiveness of audit trail traceability in practice, and tooling issues of software development and regulatory activities.  

\subsubsection{Traceability audit trail effectiveness}

Our approach enables compliance officers to perform their activities using the same tool used by the development team. They are able to track the decomposition of the design inputs in form of labelled GitHub issues, starting with the system requirements, going through the high level software requirements, and ending with the low level or detailed software requirements. The change management is performed at every level during the pull request review phase, which serves also as a design control. During the pull request review, the regulatory activities are performed and the evidence trail is collected by building relationships between requirements, test cases and artifacts contained in the pull request, according to the traceability information model. The highly automated process, with human input limited to very specific procedures, enables rapid and continuous software certification without the need of special tools (e.g. Sherlock \cite{evidence}).

Lightweight formats familiar to developers like markdown, serve as effective means to document design inputs (e.g. issues), and design outputs (e.g. software architecture and design augmented with PlantUML\footnote{https://plantuml.com} or Mermaid\footnote{https://mermaid-js.github.io/mermaid/} diagrams). Being text-based, these design documents can be properly version either directly into GitHub, as is the case of issues, or in the git repository for all other documents. Additionally, keeping the design documents close to the code and performing the change management activities in a single step (e.g. pull request review) ensures that the documentation is properly maintained, following the software development pace.

\subsubsection{Common tooling for software development and regulatory activities}

Traditionally, ALM tools address product lifecycle management, covering governance, development, and maintenance. These include management, software architecture, programming, testing, maintenance, change management, integration, project management, and release management. However, as already mentioned, these often require manual interventions from developers, and a waterfall-like approach favored by compliance officers is often prescribed in them as the advocated process. Hence, a divide between software developers and compliance officers emerges.


Distributed teams, sophisticated version management systems, and increasing use of real-time collaboration have given rise to the practice of integrated application lifecycle management, or integrated ALM, where all the tools and tools' users are synchronized with each other throughout the application development stages. The proposed tool falls to this category, building on these capabilities that are immediately available in GitHub and on an extensions that support tracing the artifacts needed for compliance reasons. This in essence integrates regulatory activities in the continuous software engineering pipeline. This in particular concerns pull requests, which are the way to introduce changes to software, but which can also be used as means to manage compliance with respect to changes in code.

The proposed implementation is at present only at prototype stage. However, although the approach looks rough comparing with the much more polished ALM tools, it has several benefits that can be associated with the use of state-of-the-art software engineering tools and associated ecosystems. These include (i) leveraging a large 3rd party DevOps tools ecosystem, which includes numerous beneficial tools and subsystems that are available either in open source or as hosted services; (ii) the solid GitHub APIs, which are used in numerous GitHub projects; and (iii) close integration with popular development environments such as Visual Studio Code\footnote{https://code.visualstudio.com}.


\subsubsection{Limitations and future work}

The effective use of the proposed approach requires a level of familiarity with GitHub and related the DevOps ecosystem. Although this should be the case for experienced software-intensive organizations, traditional medical device manufacturers and compliance professionals may find it difficult to switch from an integrated document oriented compliance process to one where the documentation is managed as code and the authoring tools are not word processors or spreadsheet applications. Better authoring tools and simpler ways of navigating the GitHub user interface for non-programmers would simplify the adoption process and make this way of working more accessible.

\section{Conclusions}

Developing regulated software is often considered as an activity that is complicated by compliance related aspects, such as traceability and risk management. For many organizations, this has meant using waterfall-like development approaches, where the sequential phases help in managing traceability. However, such approach in essence eliminates the opportunity to use agile or continuous software engineering methods.

To improve the situation, in this paper we have described our approach that expands the GitHub functionality with traceability from requirements to implementation, a key element when developing safety critical software systems. Our prototype implementation demonstrates that GitHub serves as an effective design control mechanism, allowing regulatory professionals to conduct their regulatory activities alongside software developers.

\section*{Acknowledgements}
The authors would like to thank Business Finland and the members of the AHMED (Agile and Holistic MEdical software Development) consortium for their contribution in preparing this paper.

%
%

\bibliographystyle{splncs04}
\bibliography{references}

\end{document}